\documentclass{PoS}

\newcommand{\hspn}{{\hspace{-4.5mm}}}
\newcommand{\beq}{\begin{equation}}
\newcommand{\eeq}{\end{equation}}
\newcommand{\bea}{\begin{eqnarray}}
\newcommand{\eea}{\end{eqnarray}}
\newcommand{\nn}{\nonumber}

\newcommand{\as}{\alpha_{\rm s}}

\newcommand{\ar}{a_{\rm s}}
 
\def\z#1{{\zeta_{#1}^{}}}
\def\ca{{C^{}_A}}
\def\cf{{C^{}_F}}
\def\nf{{n^{}_{\! f}}}

\def\cfs{{C_{F}^{\,2}}}
\def\cfa{{C_{FA}}}
\def\cfas{{C_{FA}^{\,2}}}

\def\zs#1{{\zeta_{#1}^{\,2}}}
\def\zt#1{{\zeta_{#1}^{\,3}}}

\def\pqq(#1){p_{\rm{qq}}(#1)}
\def\H(#1){{\rm{H}}_{#1}^{}}
\def\Hh(#1,#2){{\rm{H}}_{#1,#2}^{}}
\def\Hhh(#1,#2,#3){{\rm{H}}_{#1,#2,#3}^{}}
\def\Hhhh(#1,#2,#3,#4){{\rm{H}}_{#1,#2,#3,#4}^{}}

\title{CC DIS at $\alpha_s^3$ in Mellin-$N$ and Bjorken-$x$ spaces}

\ShortTitle{CC DIS at $\alpha_s^3$ in Mellin-$N$ and Bjorken-$x$ spaces}

\author{\speaker{M.~Rogal}
\\
        Deutsches Elektronensynchrotron DESY \\
        Platanenallee 6, D--15738 Zeuthen, Germany\\
        E-mail: \email{Mikhail.Rogal@desy.de}}

\abstract{Third-order results for the structure functions of
charged-current deep-inelastic scattering are discussed. New results for 11'th Mellin moment for  $\,F_{\:\!2,L}^{\,\nu p - \bar\nu p}$ structure functions and 12'th moment for  $\,F_{\:\!3}^{\,\nu p - \bar\nu p}$ are presented as well as corresponding higher Mellin moments of
differences between the respective crossing-even 
and -odd coefficient functions.
Approximations in Bjorken-$x$ space for these differences obtained with lowest five moments  as well as consistency of new results with these approximations are discussed.  
The $1/N_c$ suppression of the differences is shown and the correction to the Paschos-Wolfenstein relation is discussed.}

\FullConference{8th International Symposium on Radiative Corrections (RADCOR)\\
		 October 1-5 2007\\
		 Florence, Italy}

\begin{document}

\section{Introduction}
Structure functions in deep-inelastic scattering (DIS) are among the most 
extensively measured observables. Today the combined data from fixed-target 
experiments and the HERA collider spans about four orders of magnitude in both
Bjorken-$x$ and the scale $Q^2 = -q^2$ given by the momentum $q$ of the 
exchanged electroweak gauge boson \cite{Yao:2006px}. 
In this report I focus on the $W\!$-exchange charged-current (CC) case, see
Refs.~\cite {Tzanov:2005kr,Aktas:2005ju,Chekanov:2006da} for recent 
measurements in neutrino DIS and at HERA. I present new Mellin moments for coefficient functions in combination $\nu p - \bar\nu p$ and discuss the results for differences between the corresponding crossing-even 
and -odd coefficient functions.
I show suppression of such differences in large $1/N_c$ limit and discuss $\alpha_s^3$ correction to the Paschos-Wolfenstein relation~\cite{Paschos:1973kj}. 
\section{Results for the CC coefficient functions and their applications}
\label{sec:results}
%
%
Recently the first five odd-integer moments have been computed of the third-%
order coefficient functions for $\,F_{\:\!2,L}^{\,\nu p - \bar\nu p}$ in 
charged-current DIS, together with the corresponding moments 
$\,N = 2,\, \ldots ,\, 10\,$ for $\,F_{\:\!3}^{\,\nu p - \bar\nu p\!}$ 
\cite{Moch:2007gx}. 
Meanwhile  calculation of new results for 11'th moment for the first and 12'th moment for the latter case has been finished. We use scale choice $\,\mu_r = \mu_f^{} = Q\,$ and standard QCD colour factors $\ca=3$ and $\cf=4/3$ throughout this paper and denote the Mellin-$N$ moments of corresponding coefficient functions as $C_{a,N}^{\rm ns},\, a=2,3,L$. Following the formalism outlined in  \cite{Moch:2007gx} 
we find the following numerical results
\begin{eqnarray}
\label{eq:c2ns11}
 C_{2,11}^{\rm ns} & = &
        1+ 21.01295976\, \ar 
         + \ar^2\,  (
            722.3767644
          - 51.01867375 \, \nf
          )
\nonumber\\
& &\mbox{}
       + \ar^3\,  (
            29020.51723
          -4259.717409\, \nf
          + 89.53420655\, \nf^2
          )
 \nonumber\, ,
\quad
\\
 C_{L,11}^{\rm ns} & = &
       0.4444444444\, \ar 
         + \ar^2\,  (
            30.42631299
          - 1.781422693 \, \nf
          )
\nonumber\\
& &\mbox{}
       + \ar^3\,  (
            2021.685213
          - 266.3750306 \, \nf
          + 7.082458684\, \nf^2
          )   
 \nonumber\, ,
\quad
\\
\label{eq:c3ns12}
 C_{3,12}^{\rm ns} & = &
      1+  22.20106054 \ar 
         + \ar^2\,  (
            774.6238566
          - 53.26617873\, \nf
          )
\nonumber\\
& &\mbox{}
       + \ar^3\,  (
            31152.95983
          - 4483.444700 \, \nf
          + 91.41515482 \, \nf^2
          )     
\, ,
\end{eqnarray}
where the normalized coupling constant $\ar = \as /(4 \pi)$ and  $\nf$ denotes the number of effectively massless quark flavours. The results in analytical form can be found in App.~\ref{app:cc-mom}.

Unlike  fixed-$N$ calculations for the combination $\nu p - \bar\nu p$, the complete
three-loop results for $F_{\:\!2,L}^{\,\nu p + \bar \nu p\,}$ 
\cite{Moch:2004xu,Vermaseren:2005qc} (the $\as^3$ coefficient functions for this process are those of 
 photon-exchange DIS, but without the contributions of the $fl_{11}$ flavour
 classes)
and $F_{\:\!3}^{\nu P + \bar \nu P}$ \cite{Vogt:2006bt} facilitate analytic 
continuations to these values of $N$. This continuation has been performed  
using the $x$-space expressions 
and the Mellin transformation package provided with version 3 
of {\sc Form}~\cite{Vermaseren:2000nd}.  
Thus we are in a position to derive the respective moments of
the hitherto unknown third-order contributions to the even-odd differences which are defined as
\bea
  \label{eq:cdiff}
  \delta\, C_{2,L} \; =\; C_{2,L}^{\,\nu p + {\bar \nu} p} 
    - C_{2,L}^{\,\nu p - {\bar \nu} p} \:\: , \qquad
  \delta\, C_3 \; =\;  C_3^{\,\nu p - {\bar \nu} p}
    - C_3^{\,\nu p + {\bar \nu} p} \:\: .
\eea
The signs are chosen such that the differences are always `even -- odd' in the 
moments $\, N$ accessible by the OPE, and it is understood 
that the $d^{\:\!abc}d_{abc}$ part of $\,C_3^{\,\nu p + \bar\nu p}$ 
\cite{Retey:2000nq,Vogt:2006bt} is removed before the difference is formed. The
non-singlet quantities (\ref{eq:cdiff}) have an expansion in powers of $\as$, 
\bea
\label{eq:cf-exp}
  \delta\, C_a \; = \; 
  \sum_{l=2} \: \ar^{\, l}\: \delta\:\! c_{a}^{(l)} .
\eea
There are no first-order contributions to these differences, hence the sums
start at $l = 2\,$ in Eq.~(\ref{eq:cf-exp}).
Here I present only numerical results for the differences corresponding to the Mellin moments in Eqs.~(\ref{eq:c2ns11})
whereas the corresponding results in analytical form are shown in App.~\ref{app:MM-diff}.
The lower Mellin moments of such differences can be found in Ref.~\cite{MRV1}.
Using the notation $\delta\, C_{a,\,N}$ for the $N$-th moment of
$\delta\, C_{a}(x)$
the results for higher Mellin moments read 
\bea
\label{eq:dc2ns11}
  \delta\, C_{2,11} & = &
          - 0.004083868756
            \, \ar^2
        \: + \:\ar^3 \,  \*  (
       + 0.1559414787
          -0.01710053059  \* \, \nf
          )\:\: ,
\nonumber
\\
  \delta\, C_{L,11} & = &
          -0.001670175019
          \, \ar^2
       \: + \:\ar^3 \,  \*  (
          - 0.3317043993
          +  0.006939009889\* \, \nf
          )\:\: , 
\nonumber
\\
  \delta\, C_{3,12} & = &
          -0.009709081656
           \, \ar^2
       \: + \:\ar^3 \,  \*  (
          -0.6201718804
          +0.01191844785 \* \, \nf
          ) \:\: .
\eea
The new $\as^3$ contributions are rather large if compared to the leading
second-order results also included in Eqs.~(\ref{eq:dc2ns11})
with, e.g., $\,\ar = 1/50\,$ corresponding to $\,\as \simeq 0.25$. On the other hand, the integer-$N$ 
differences $\delta\, C_{a,N}$ are entirely negligible compared to the 
$\,\nu p \pm \bar\nu p\,$ moments $C_{a,N}$ of Eqs.~(\ref{eq:c2ns11}) and Refs.~\cite{Retey:2000nq,Moch:2007gx}.

To discuss the colour structure of the results I present  the $\alpha_s^3$ part for  $\delta\, C_{2,11}$ 
with analytical colour factor dependence
\begin{equation}
 \delta c_{2,11}^{(3)}= 
 0.9495866025 \, \cf \* \cfas
 -0.4076041653 \, \cfs \* \cfa
 + 0.07695238768 \, \cf \* \cfa \* \nf \, .
\end{equation}
One notes here that result contains an overall factor $\cfa=\cf-\ca/2=-1/(2N_c)\,$. This occurrence is typical for all calculated  differences
$\delta c_{a,\,N}^{(3)}$. 
Hence the third-order even-odd differences are suppressed in the large-$N_c$ 
limit as conjectured, to all orders, in Refs.~\cite{Broadhurst:2004jx,Kataev:2007jz}

Let us now consider 
consequences of the moments of the type Eqs.~(\ref{eq:dc2ns11}) for the $x$-space functions 
$\delta\:\! c_{a}^{(3)}(x)$. 
For moment-based approximations a simple ansatz was chosen, and its free parameters were determined from the  first five moments  
available in Ref.~\cite{MRV1}.
This ansatz 
is then varied in order to estimate the remaining uncertainties. Finally 
two 
approximations, denoted below by $A$ and $B$, are selected which indicate the
widths of the uncertainty bands. For $F_{\:\! 2}$ 
these functions are, with $\, L_0 = \ln x\,$, 
$x_1 = 1-x$  and $\, L_1 = \ln x_1$, 
\bea
\label{eq:dc2qq3p}
 \delta c_{2,\,A}^{(3)}(x) &\! =\! &
   ( 54.478\,L_1^2 + 304.6\,L_1 + 691.68\, x ) \, x_1
   + 179.14\,L_0 - 0.1826\,L_0^3
 \nn \\ & & \mbox{\hspn}
 + \nf\, \{ ( 20.822\, x^2 - 282.1\, (1 + {\textstyle {x \over 2}}) )\, x_1
   - (285.58\, x + 112.3 - 3.587\,L_0^2) \, L_0 \}
 \:\: , \nn \\[0.5mm]
 \delta c_{2,\,B}^{(3)}(x) &\! =\! &
   - ( 13.378\,L_1^2 + 97.60\,L_1 + 118.12\, x ) \, x_1
   - 91.196\,L_0^2 - 0.4644\,L_0^5  
\nn \\ & & \mbox{\hspn}
 + \nf\, \{ (4.522\,L_1 + 447.88\, (1 + {\textstyle {x \over 2}}) ) \, x_1
   + (514.02\, x + 147.05 + 7.386\,L_0)\, L_0 \}  \:\: .
\eea
The uncertainty band presented by Eqs.~(\ref{eq:dc2qq3p}) does not directly indicate the range of applicability, since the coefficient functions enter observables only via smoothening Mellin convolutions with non-perturbative 
initial distributions. As result theoretical uncertainties for physical observables become even smaller (see Ref.~\cite{MRV1} for details). 
It is worth to mention that the new Mellin moments in Eqs.~(\ref{eq:c2ns11}) and (\ref{eq:dc2ns11}) are consistent with the  approximations based on the first five Mellin moments only, thus confirming the reliability of the uncertainty estimates.

The approximations~(\ref{eq:dc2qq3p}) and analogue of it for $\delta c_{L}^{(3)}(x)$ can be used to determine $\alpha_s^3$ corrections to the Paschos-Wolfenstein relations (see  Ref.~\cite{MRV1} and references therein for details and discussion) 
\bea
\label{eq:rminus-numbers}
R^{-} &\! = &
 \frac{1}{2} - \sin^2\theta_W 
  \:\: + \:\: \frac{u^- - d^- + c^- - s^-}{u^- + d^-} \: \Bigg\{
  1 - \frac{7}{3}\:\sin^2\theta_W  
  \; + \; \left( \frac{1}{2} - \sin^2\theta_W \right) \cdot
\nn \\ & & \mbox{} 
  \frac{8}{9\,\pi}\left[ \,
    \as
  + 1.689\,\as^2
  + (3.661 \pm 0.002)\,\as^3 \,
  \right]
  \Biggr\}
  \; + \; {\cal{O}} \left( (u^- + d^-)^{-2\,} \right)
  \; + \; {\cal{O}}\bigl(\alpha_s^4\bigr) 
\:\: . \quad
\eea
The ratio~(\ref{eq:rminus-numbers}) is an expansion in $\alpha_s$ and inverse powers of the dominant isoscalar combination $u^- + d^-$, where $ q^- =\int_0^1 dx\; x \left( q(x) - \bar{q}(x) \right)$ is the second Mellin moment of the valence quark distributions. The third term in the square brackets is determined with our $\alpha_s^3$ corrections and the perturbation series  appears reasonably well 
convergent 
although the correction in not negligible.
On the 
other hand, due to the small prefactor of this expansion, the new third-order 
term increases the complete curved bracket in Eq.~(\ref{eq:rminus-numbers}) by only 
about a percent, which can therefore by considered as the new uncertainty of this 
quantity due to the truncation of the perturbative expansion.
\section{Summary}
\label{sec:summary}
In this report I have presented new results for 11'th Mellin moment for  $C_{2}^{\rm ns}$
 and $C_{L}^{\rm ns}$ 
Wilson coefficient functions  as well as 12'th moment for $C_{3}^{\rm ns}$ function. I have discussed the Mellin moments for differences between the corresponding crossing-even 
and -odd coefficient functions and use of these to obtain approximations in Bjorken-$x$ space which are ready for phenomenology applications (see, e.g.,  Ref.~\cite{Alekhin:2007fh}). It was shown that the differences are suppressed by the number of colours.  The third order QCD correction to the Paschos-Wolfenstein relation was obtained 
with help of the approximations in Bjorken-$x$ space. The correction was found to be small.

{\sc Form} file of the results (App.~\ref{app:cc-mom} and App.~\ref{app:MM-diff}) can be obtained from the preprint server
{\tt http://arXiv.org} by downloading the source of this article.

\acknowledgments
I would like to thank S.~Moch for fruitful discussions and proofreading. This paper was finalized at the Galileo Galilei Institute for Theoretical Physics during  the workshop ``Advancing Collider Physics: from Twistors to Monte Carlos''.
\appendix
\section{Appendix}
\label{app:cc-mom}
In this Appendix I present the analytic expressions up to order $\ar^3$ 
for the coefficient functions 
$C_{2,11}^{\rm ns}$, $C_{L,11}^{\rm ns}$  
and $C_{3,12}^{\rm ns}$ at
the scale $\mu_r = \mu_f = Q$. 
The notation follows Sec.~\ref{sec:results} where these functions were presented numerically in Eqs.~(\ref{eq:c2ns11}). $C_A$ and $C_F$ are
the standard QCD colour factors, 
$C_A \equiv N_c$ and $C_F = (N_c^2 -1)/(2N_c)$, and 
$\zeta_i$ stands for Riemann's $\zeta$-function. 
The Wilson coefficients read
\begin{eqnarray}
\label{eq:c2q11}
  C_{2,11}^{\rm ns} & = &
1
\nonumber\\
& &\mbox{}
       + \ar \* \cf \* {13105783\over 831600}
\nonumber\\
& &\mbox{}
       + \ar^2 \* \cf \* \nf  \*  \Biggl(
          - {122253517912789\over 3195000547200}
          \Biggr)
\nonumber\\
& &\mbox{}
       + \ar^2 \* \cf^2  \*  \Biggl(
          {98146880716389133\over 1845112816008000} + {8462\over 105} \* \z3
          \Biggr)
\nonumber\\
& &\mbox{}
       + \ar^2 \* \ca \* \cf  \*  \Biggl(
           { 236212260301543\over 1037337840000} - {109357\over 1155} \* \z3
          \Biggr)
\nonumber\\
& &\mbox{}
       + \ar^3 \* \cf \* \nf^2  \*  \Biggl(
           {63689629686066726367\over 996360920644320000} + {251264\over 
         93555} \* \z3 
          \Biggr)
\nonumber\\
& &\mbox{}
       + \ar^3 \* \cf^2 \* \nf  \*  \Biggl(
          - {7118183480913604311036887\over 10880261253435974400000}
          + {125632\over 3465 }\* \z4 - {8620003991\over 85135050} \* \z3 
          \Biggr)
\nonumber\\
& &\mbox{}
       + \ar^3 \* \cf^3  \*  \Biggl(
           -  { 35440239155810867032199497\over 3544454339100103968000000}
          - {81164\over 99 }\* \z5 + {151689577\over 8004150} \* \z4
\nonumber\\
& &\mbox{}{\hspn}\qquad\qquad\quad
 + {381164463904607\over 249609417750} \* 
         \z3 
          \Biggr)
\nonumber\\
& &\mbox{}
       + \ar^3 \* \ca \* \cf \* \nf  \*  \Biggl(
          - {206484440760943457173\over 203899125830400000} - {125632\over 
         3465} \* \z4 + {244949367479\over 936485550} \* \z3 
          \Biggr)
\nonumber\\
& &\mbox{}
       + \ar^3 \* \ca \* \cf^2  \*  \Biggl(
          {130339510428192960678390973\over 54401306267179872000000}
          - {21218\over 99} \* \z5 - {151689577\over 5336100 }\* \z4 
\nonumber\\
& &\mbox{}{\hspn}\qquad\qquad\quad
+ {80821712466167\over 576875098800 }\* 
         \z3 
          \Biggr)
\nonumber\\
& &\mbox{}
       + \ar^3 \* \ca^2 \* \cf  \*  \Biggl(
           {6214478354002179611868649\over 1884027922672896000000} + 
         {197732\over 231} \* \z5 + {151689577\over 16008300} \* \z4 
\nonumber\\
& &\mbox{}{\hspn}\qquad\qquad\quad
- {1039437594401\over 416215800} \* \z3 
          \Biggr)
\, ,
\quad
\\
\label{eq:cLq11}
  C_{L,11}^{\rm ns} & = &
      \ar \* \cf  \*  
         { 1\over 3} 
\nonumber\\
& &\mbox{}
       + \ar^2 \* \cf \* \nf  \*  \Biggl(
           - {15151\over 11340}
          \Biggr)
\nonumber\\
& &\mbox{}
       + \ar^2 \* \cf^2  \*  \Biggl(
          - {313264177\over 104781600} + 8 \* \z3 
          \Biggr)
\nonumber\\
& &\mbox{}
       + \ar^2 \* \ca \* \cf  \*  \Biggl(
          {1984474543\over 209563200} - 4 \* \z3  
          \Biggr)
\nonumber\\
& &\mbox{}
       + \ar^3 \* \cf \* \nf^2  \*  
          {1355317\over 255150}
\nonumber\\
& &\mbox{}
       + \ar^3 \* \cf^2 \* \nf  \*  \Biggl(
         {355389053714327\over 8307001422720} - {30048196\over 405405} \* \z3
          \Biggr)
\nonumber\\
& &\mbox{}
       + \ar^3 \* \cf^3  \*  \Biggl(
          - {9102853788548773397\over 27676692240120000} + {4240\over 3} \* \z5 - 
         {505404892\over 571725} \* \z3
          \Biggr)
\nonumber\\
& &\mbox{}
       + \ar^3 \* \ca \* \cf \* \nf  \*  \Biggl(
           - {119231997944341913\over 1213685272800000} + {1849566364\over 42567525} \* \z3
          \Biggr)
 \nonumber\\
& &\mbox{}
      + \ar^3 \* \ca \* \cf^2  \*  \Biggl(
          - {5696808539467201991\over 290745049795200000} - 1480 \* \z5
          + {1327717221659\over 936485550} \* \z3 
          \Biggr)
\nonumber\\
& &\mbox{}
       + \ar^3 \* \ca^2 \* \cf  \*  \Biggl(
         {278571247117091333\over 882680198400000} + {1160\over 3} \* \z5 - 
         {89146587169\over 170270100} \* \z3
          \Biggr)
\, ,
\quad
\\
\label{eq:c3q12}
  C_{3,12}^{\rm ns} & = &
 \nonumber      1 
\\
& &\mbox{}
       + \ar \* \cf  \*  
          {180008419\over 10810800 }
\nonumber\\
& &\mbox{}
       + \ar^2 \* \cf \* \nf  \*  \Biggl(
          - {56084621695749173\over 1403883240439680}
          \Biggr)
\nonumber\\
& &\mbox{}
       + \ar^2 \* \cf^2  \*  \Biggl(
          {3728023695177860551381\over 52698267138004488000} + {88882\over 1155}
          \* \z3
          \Biggr)
 \nonumber\\
& &\mbox{}
      + \ar^2 \* \ca \* \cf  \*  \Biggl(
        { 41265384392827325207\over 175485405054960000} - {1424581\over 15015}
          \* \z3
          \Biggr)
 \nonumber\\
& &\mbox{}
      + \ar^3 \* \cf \* \nf^2  \*  \Biggl(
           {1855782116991095763359887\over 28457064254522423520000} + 
         {3387392\over 1216215} \* \z3
          \Biggr)
\nonumber\\
& &\mbox{}
       + \ar^3 \* \cf^2 \* \nf  \*  \Biggl(
          - {22132972736542162287370477597\over 
         28250103787216805894400000} + {1693696\over 45045 }\* \z4 - {496067497273\over 
         12174312150} \* \z3 
         \Biggr)
\nonumber\\
& &\mbox{}
       + \ar^3 \* \cf^3  \*  \Biggl(
          - {1556243722103827358575546080538633\over 
         17108404104057433733678112000000} -{ 551372\over 1287} \* \z5 + {25648239313\over 
         1352701350 }\* \z4 
\nonumber\\
& &\mbox{}{\hspn}\qquad\qquad\quad
+ {67198327643555959\over 49853808254250} \* \z3
          \Biggr)
\nonumber\\
& &\mbox{}
       + \ar^3 \* \ca \* \cf \* \nf  \*  \Biggl(
          - {383732229667234059139588387\over 
         379427523393632313600000 }-{ 1693696\over 45045} \* \z4 + {31504273327\over 133783650} \* 
         \z3  
          \Biggr)
\nonumber\\
& &\mbox{}
       + \ar^3 \* \ca \* \cf^2  \*  \Biggl(
           {55562394749867369342925789360923\over 
         17091312791266167566112000000} - {58298\over 99} \* \z5 -{ 25648239313\over 901800900} \*       \z4
\nonumber\\
& &\mbox{}{\hspn}\qquad\qquad\quad\quad
 - {518595205781183\over 6499459446480 }\* \z3 
           \Biggr)
\nonumber\\
& &\mbox{}
       + \ar^3 \* \ca^2 \* \cf  \*  \Biggl(
           {145036477650625443670858543223\over 
         45531302807235877632000000} + {2899516\over 3003} \* \z5 + {25648239313\over 2705402700}          \* \z4 
\nonumber\\
& &\mbox{}{\hspn}\qquad\qquad\quad\quad
- {118955981330663\over 48697248600} \* \z3
          \Biggr)
\, .
\end{eqnarray}

\section{Appendix}
\label{app:MM-diff}

Here I present the analytic expressions for the Mellin-space 
coefficient-function differences $\delta c_{2,11}^{(3)}$ $\delta c_{L,11}^{(3)}$ and $\delta c_{3,12}^{(3)}$ 
given 
numerically in Eqs.~(\ref{eq:dc2ns11}). We use the 
notations and conventions as specified in 
Sec.~\ref{sec:results}.
The moments of this quantity are given by
\bea
\label{eq:dc2q11}
  \delta c_{2,11}^{(3)} &\! =\! &
        \cf \* \cfas  \*  \Biggl(
            {72985545040605109471734789941\over 12566701747718550432000000}
          - {23832\over 77} \* \z5
          + {55689927719519927\over 1622461215375} \* \z3
\nonumber\\
& &\mbox{\hspn}\qquad\qquad
          - {1472\over 3} \* \zs3
          - {15177966246339422387\over 1537594013340000} \* \z2
          + {8654312\over 945} \* \z2 \* \z3
 - {31781239759\over 1819125} \* \zs2
\nonumber\\
& &\mbox{\hspn}\qquad\qquad
          - {8992\over 63} \* \zt2
          \Biggr)
\nonumber\\
& &\mbox{\hspn}
       + \cfs \* \cfa  \*  \Biggl(
           - {417272486089283995425649952101\over 69116859612452027376000000}
          + {66993964\over 10395} \* \z5
 \nonumber\\
& &\mbox{\hspn}\qquad\qquad
          - {436861012176654887\over 12979689723000} \* \z3
          + {1456\over 3} \* \zs3
  + {1514141807453257771\over 109828143810000} \* \z2
\nonumber\\
& &\mbox{\hspn}\qquad\qquad
          - {77266484\over 10395} \* \z2 \* \z3
          + {2125039226077\over 180093375 }\* \zs2
          - {56432\over 315} \* \zt2
          \Biggr)
\nonumber\\
& &\mbox{\hspn}
       + \nf \* \cf \* \cfa \*  \Biggl(
           {188706965915502835794319\over 369391585764801600000}
          - {1792\over 9} \* \z5
          + {22120534934\over 10405395} \* \z3
 \nonumber\\
 & &\mbox{}\qquad\qquad
          - {3063427279802429\over 2995313013000} \* \z2
          + {1408\over 9} \* \z2 \* \z3
          - {9519404\over 17325} \* \zs2
          \Biggr)
\; ,
\quad
\\[2mm]
\label{eq:dcLq11}
  \delta c_{L,11}^{(3)} & = &
        \cf \* \cfas  \*  \Biggl(
         {13469264008826648669111\over 9251922834554400000}
          - {8288\over 3 }\* \z5
          + {10937259706\over 4729725} \* \z3
\nonumber\\
& &\mbox{\hspn}\qquad\qquad
          - {12442705009487\over 6483361500} \* \z2
          + 1312 \* \z2 \* \z3
          - {1413232\over 4725} \* \zs2
          \Biggr)
\nonumber\\
& &\mbox{\hspn}
       + \cfs \* \cfa  \*  \Biggl(
          {29154136441960450330061\over 226672109446582800000}
          - {368\over 3} \* \z5
          - {560828132092\over 468242775} \* \z3
\nonumber\\
& &\mbox{\hspn}\qquad\qquad
          + {244988263837\over 480249000} \* \z2
          + 48 \* \z2 \* \z3
          + {9705316\over 51975} \* \zs2
          \Biggr)
\nonumber\\
& &\mbox{\hspn}
       + \nf \* \cf \* \cfa  \*  \Biggl(
           {717577454838293479\over 32708818101960000}
          + {26725912\over 405405} \* \z3
          - {67470181\over 1403325} \* \z2
          - {368\over 45 }\* \zs2
          \Biggr)
\; ,
\quad
\\[2mm]
\label{eq:dc3q12}
  \delta c_{3,12}^{(3)} & = &
       \cf \* \cfas  \*  \Biggl(
 - {29715680590481634053309842446559253\over 
         4665928392015663745548576000000}
          + {2637896\over 3003 }\* \z5
\nonumber\\
& &\mbox{\hspn}\qquad\qquad
          - {9720046002002881973\over 274195945398375} \* \z3
          + {1472\over 3} \* \zs3
+{151219518094861415132729\over 14638407538334580000} \* \z2
 \nonumber\\
& &\mbox{\hspn}\qquad\qquad
          - {1290033016\over 135135} \* \z2 \* \z3
          + {20250388084007\over 1127251125} \* \zs2
          + {8992\over 63} \* \zt2  
          \Biggr)
\nonumber\\
& &\mbox{\hspn}
       + \cfs \* \cfa  \*  \Biggl(
           {165791456833171990576039064477409253\over 
         25662606156086150600517168000000}
          - {902648332\over 135135 }\* \z5
\nonumber\\
& &\mbox{\hspn}\qquad\qquad
          + {6932482134993002293\over 199415233017000} \* \z3
          - {1456\over 3} \* \zs3
          - {3478984519432985388157\over 241292431950570000} \* \z2
\nonumber\\
& &\mbox{\hspn}\qquad\qquad
          + {79428644\over 10395 }\* \z2 \* \z3
          - {369039205594393\over 30435780375 }\* \zs2
          + {56432\over 315} \* \zt2
          \Biggr)
\nonumber\\
& &\mbox{\hspn}
       + \nf \* \cf \* \cfa  \*  \Biggl(
          - {1297953990099826399951147609\over 2434659941775807345600000}
          + {1792\over 9 }\* \z5
          - {876359298658\over 405810405} \* \z3
\nonumber\\
& &\mbox{\hspn}\qquad\qquad
          +{ 40974440556905057\over 38939069169000} \* \z2
          - {1408\over 9} \* \z2 \* \z3
          + {13881228\over 25025}\* \zs2
          \Biggr)
\; .
\end{eqnarray}


\begin{thebibliography}{99}

\bibitem{Yao:2006px}
{\bf Particle Data Group} Collaboration, W.~M. Yao {\em et~al.}, 
{\em J. Phys.} {\bf G33} (2006) 1--1232.

\bibitem{Tzanov:2005kr}
{\bf NuTeV} Collaboration, M.~Tzanov {\em et~al.}, 
 {\em Phys. Rev.}
  {\bf D74} (2006) 012008  [\href{http://xxx.lanl.gov/abs/hep-ex/0509010}{{\tt
  hep-ex/0509010}}].

\bibitem{Aktas:2005ju}
{\bf H1} Collaboration, A.~Aktas {\em et~al.},  
  {\em Phys. Lett.} {\bf B634} (2006) 173--179 
  [\href{http://xxx.lanl.gov/abs/hep-ex/0512060}{{\tt hep-ex/0512060}}].

\bibitem{Chekanov:2006da}
{\bf ZEUS} Collaboration, S.~Chekanov {\em et~al.}, 
  {\em Phys. Lett.} {\bf B637} (2006)
  210--222 [\href{http://xxx.lanl.gov/abs/hep-ex/0602026}{{\tt
  hep-ex/0602026}}].

\bibitem{Paschos:1973kj}
E.~A. Paschos and L.~Wolfenstein, 
 {\em Phys. Rev.} {\bf D7} (1973) 91--95.

\bibitem{Moch:2007gx}
S.~Moch and M.~Rogal, 
  {\em Nucl. Phys.} {\bf B782} (2007) 51--78
  [\href{http://xxx.lanl.gov/abs/arXiv:0704.1740 [hep-ph]}{{\tt arXiv:0704.1740
  [hep-ph]}}].

\bibitem{Moch:2004xu}
S.~Moch, J.~A.~M. Vermaseren and A.~Vogt, 
  {\em Phys. Lett.} {\bf B606} (2005) 123--129
  [\href{http://xxx.lanl.gov/abs/hep-ph/0411112}{{\tt hep-ph/0411112}}].

\bibitem{Vermaseren:2005qc}
J.~A.~M. Vermaseren, A.~Vogt and S.~Moch, 
 {\em Nucl. Phys.} {\bf
  B724} (2005) 3--182 [\href{http://xxx.lanl.gov/abs/hep-ph/0504242}{{\tt
  hep-ph/0504242}}].

\bibitem{Vogt:2006bt}
A.~Vogt, S.~Moch and J.~Vermaseren, 
{\em Nucl. Phys. Proc. Suppl.} {\bf 160}
  (2006) 44--50 [\href{http://xxx.lanl.gov/abs/hep-ph/0608307}{{\tt
  hep-ph/0608307}}].

\bibitem{Vermaseren:2000nd}
J.~A.~M. Vermaseren, 
 {\em Preprint}  (2000)
  [{{\tt math-ph/0010025 }}].

\bibitem{Retey:2000nq}
A.~Retey and J.~A.~M. Vermaseren, 
  {\em Nucl. Phys.} {\bf B604} (2001) 281--311
  [\href{http://xxx.lanl.gov/abs/hep-ph/0007294}{{\tt hep-ph/0007294}}].

\bibitem{MRV1}
S.~Moch, M.~Rogal and A.~Vogt, 
 {\em Nucl. Phys.} {\bf B} (2007) in Press
  [\href{http://xxx.lanl.gov/abs/arXiv:0708.3731 [hep-ph]}{{\tt arXiv:0708.3731
  [hep-ph]}}].

\bibitem{Broadhurst:2004jx}
D.~J. Broadhurst, A.~L. Kataev and C.~J. Maxwell, 
  {\em Phys.
  Lett.} {\bf B590} (2004) 76--85
  [\href{http://xxx.lanl.gov/abs/hep-ph/0403037}{{\tt hep-ph/0403037}}].

\bibitem{Kataev:2007jz}
A.~L. Kataev, 
 {\em Preprint} (2007) [{{\tt arXiv:0707.2855 [hep--ph]}}].

\bibitem{Alekhin:2007fh}
S.~Alekhin, S.~A. Kulagin, and R.~Petti, 
  {\em Preprint} (2007)
 [{{\tt  arXiv:0710.0124 [hep--ph]}}].

\end{thebibliography}

\end{document}